\documentclass{mn2e}
\usepackage{graphics} 

\title[Near--IR quiescent detection of XTE\,J1118+480]{The first near infrared detection of XTE\,J1118+480 in quiescence}

\author[J.\ Miko{\l}ajewska et al.]
       {Joanna~Miko{\l}ajewska$^1$\thanks{e-mail: mikolaj@camk.edu.pl},
  Artur~Rutkowski$^1$, Denise~R.~Gon{\c c}alves$^{2,3}$ 
\newauthor
 and Anna~Szostek$^1$
        \vspace*{1mm}\\
       $^1$  N. Copernicus Astronomical Center, Bartycka  18, PL-00716 Warsaw, Poland\\
        $^2$ Instituto de Astronomia, Geof\'\i sica e Ci\^encias Atmosf\'ericas, Universidade de S\~ao  
Paulo, Rua do Mat\~ao 1226,\\ 05508-900 S\~ao Paulo, Brazil\\ 
$^3$ Instituto de Astrof\'\i sica de Canarias, E-38205 La Laguna, Tenerife, Spain }

\date{\fbox{\sc Draft Version}}

\pagerange{\pageref{firstpage}--\pageref{lastpage}}
\pubyear{2005}

\begin{document}

\maketitle

\label{firstpage}

\begin{abstract} We report the first quiescent detection of KV\,UMa, the optical counterpart of  
XTE\,J1118+480 at near infrared. The observed magnitudes and colours are consistent with a K7--M0\,V star, at the distance $1.4\pm0.2$ kpc. The light curve shows strong orbital modulation with possible contamination from a superhump detected in the quiescent optical light curves. 
 \end{abstract}


\section{Introduction}
XTE\,J1118+480 is one of the most interesting soft X-ray transients (SXT) extensively studied over whole spectral range (e.g. McClinctock et al. 2001; Wagner et al. 2001; Frontera et al. 2001). It is the first black hole binary detected in the galactic halo (e.g. Mirabel et al. 2001). It has one of the highest mass functions, $\sim 6\, \rm M_{\sun}$ (McClinctock et al. 2001; Wagner et al. 2001), and one of the lowest mass ratios, $q=M_2/M_{\rm BH} = 0.037$ (Orosz 2001).
The optical counterpart of XTE\,J1118+480, KV\,UMa has been classified as K5/7\,V star based on the optical/red spectroscopy (e.g. McClintock et al. 2003; Torres et al. 2004). 

In the beginning of 2000, the binary underwent a bright outburst lasting several months, and another outburst has started in 2005 January (e.g. Zurita et al. 2005). The optical light curves obtained during the 2000 outburst revealed the superhump periodicity of $\sim 0.17$ d, caused by precession of an eccentric accretion disk. On the other hand, photometry during late decline and near quiescence revealed characteristic ellipsoidal light variations of the secondary with the orbital period of $0.169937$ d distorted by an additional modulation interpreted as due to the superhump having a period 0.3 per cent longer than the orbital one (Zurita et al. 2002a). 
 
In this paper we report the first detection of the optical companion of XTE\,J1118+480 in the near--infrared, when the system was in the quiescent state between the 2000 and 2005 outbursts.  

\section{Observations} 

\begin{figure}
\resizebox{\hsize}{!}{\includegraphics{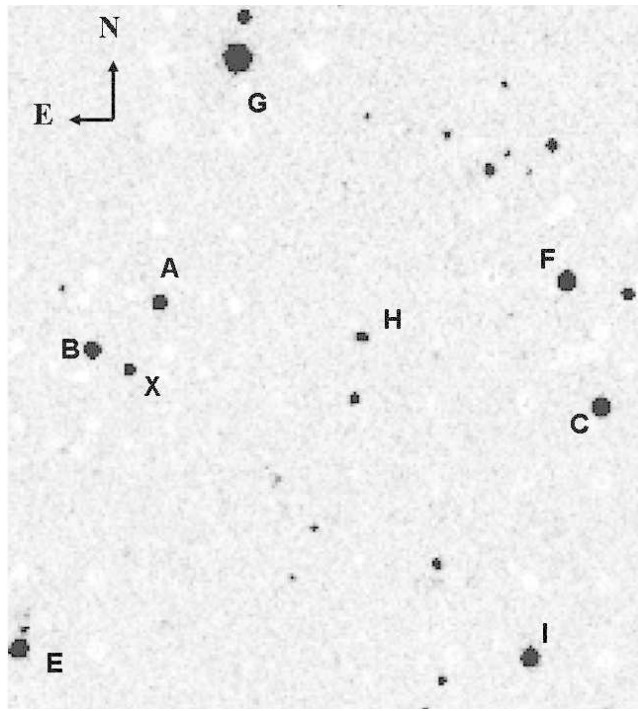}} 
\caption{The $1.3' \times 1.3'$ field of XTE\,J1118+480 (denoted by X) at $J$, composed from the 2004 March 15 observations.} \label{field} \end{figure}

XTE\,J1118+480 was observed on 2003 April 1 and 2, and 2004 March 15--18 with the CAIN II infrared camera on the 1.54 m Carlos Sanchez Telescope (TCS) at the Observatorio del Teide on Tenerife. Observations were made primarily through the $J$ filter, with an affective wavelength of 1.28 $\rm \mu m$, as well as a small number with the $K_\mathrm{s}$ ('short $K$') filter (on 2004 March 17 and 18), with an effective wavelength of 2.2 $\rm \mu m$ (Alonso, Arribas \& Martinez-Roger 1994).  Typical integration times were 60s in 2003, and 90s in 2004, respectively. The individual images were reduced in the standard way, with the data reduction performed within {\sc IRAF}\footnote{{\sc IRAF} is distributed by the National Optical Observatories, which is operated by the Association of Universities for Research in Astronomy, Inc., under contract with the NSF.}. The instrumental magnitudes were obtained using aperture photometry with the {\sc IRAF} routine {\sc APPHOT}.
For calibration we used stars in the field of XTE\,J1118+480 (Fig.~\ref{field}) selected from the 2MASS\footnote{The
Two Micron All Sky Survey (2MASS) is a
joint project of the University of Massachusetts and the Infrared Processing and Analysis
Center/California Institute of Technology, funded by the NASA and the NSF.} catalogue, and the adopted magnitudes are given in Table~\ref{field_phot}.

\section{Results and discussion}

\begin{table} 
\caption{Near infrared photometry of selected stars in the field of XTE\,J1118+480.} 
\label{field_phot}
\begin{tabular}{@{}cccccc} 
\hline \smallskip 
Star & 2MASS & $J$ & $H$ & $K$  & Ref. \cr
\hline 
A & 11180979+4802339  & 16.95 & & 16.16 & This study \cr
B & 11181198+4802190  & 16.41 &  & 15.61 & This study \cr
C & 11175562+4801591 & 15.17 & 14.41 & 14.35 & 2MASS \cr
D &                         & 17.50 &          &           & This study \cr
E & 11181437+4800433  & 15.41 & 14.83 & 14.70 & 2MASS \cr
F &11175669+4802397  & 15.03 & 14.50 & 14.27 &  2MASS \cr
G & 11180724+4803527  & 13.45 & 12.83 & 12.61 & 2MASS \cr
H &                        & 18.28 &          &    & This study \cr
I & 11175796+4803527 & 15.17 & 14.63 & 14.35 & 2MASS \cr
\hline \end{tabular} 
\end{table}

\begin{table} 
\caption{Near infrared quiescent magnitudes of XTE\,J1118+480/KV\,UMa.} 
\label{j_mag}
\begin{tabular}{@{}cccc} 
\hline \smallskip 
Date & HJD Start & HJD End & $J$   \cr
       & $+2\,450\,000$ & $+2\,450\,000$ & [mag] \cr
\hline 
~1.04.2003 & 2731.3969 & 2731.4789 & $18.09 \pm 0.03$  \cr
~2.04.2003 &2732.4295 &2732.6753& $18.08 \pm 0.03$ \cr
15.03.2004 & 3080.4728 &3080.7001& $18.01 \pm 0.02$  \cr
16.03.2004 &3081.5437 & 3081.7040 & $17.98 \pm 0.04$ \cr
17.03.2004 & 3082.4791 & 3082.6758 & $17.95 \pm 0.02$  \cr
18.03.2004 & 3083.4846 & 3083.6603 & $18.06 \pm 0.02$ \cr
\hline \end{tabular} 
\end{table}

\subsection{Near infared magnitudes, colours, and light curves}

\begin{figure}
\resizebox{\hsize}{!}{\includegraphics{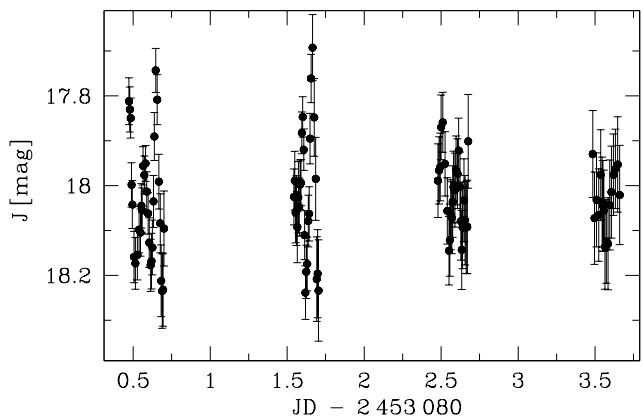}}
\caption{The $J$ light curves of KV\,UMa (XTE\,J1118+480) obtained in 2004 March 15--18. Each point represents a running mean of 3 individual measurements.}
\label{lcmar04} \end{figure}

The average $J$ magnitudes for each night are given in Table~\ref{j_mag}, and our light curves, obtained in  2004 March 15--18, are plotted in Fig.~\ref{lcmar04}.
Whereas in the $J$ filter the source was detected on a few $\sigma$ level even on a single frame, a detection in $K$ filter was possible only after addition of all the frames, and the resulting magnitude was $K=16.87 \pm 0.17$.

The average $J-K_\mathrm{s} = 1.1 \pm 0.2$ is consistent with an M--type secondary although it is redder than expected for a late K or early M--type main sequence star (Leggeit 1992; Straizys 1992, and references therein), which may indicate the presence of an evolved secondary. Similarly, our near--IR magnitudes combined with the quiescent optical magnitudes of the secondary, $V_{\rm sec} = 20.83$ and $R_{\rm sec} = 19.72$ (Wagner et al. 2001), are consistent with a  K/M secondary. In particular,  $V-J = 2.83$ and $V-K \sim 3.9$ indicate an old disc or halo M0/1 and M1\,V  star, respectively, whereas $R-J=1.72$ and $R-K=2.8$ are consistent with a K7\,V and M0\,V star, respectively (Leggeit 1992). All this suggests that the near--IR magnitudes are dominated by the secondary's light. 

The spectral type later than that derived from the optical/red spectra may be due to either the relative weakness of the absorption features in KV\,UMa as compared with the K-M dwarf templates owing to very low metallicity (e.g. Frontera et al. 2001) and/or due to an overestimated contribution of the accretion disc light.
In fact, recent estimates of the quiescent average optical magnitudes, $V=19.6$ and $R=19$ Orosz (2005), as well as $R=18.93$ (Zurita et al. 2002b), assuming  the relative contribution of the secondary $\sim 45\, \%$  the $V$ (Mc Clintock et al. 2003) and $\sim 55\, \%$ to $R$ light  (Torres et al. 2004) give the secondary magnitudes, $V_{\rm sec} \sim 20.5$ and $R_{\rm sec} \sim 19.6$--$19.7$, and optical/near--IR colours consistent with an old disc/halo K7/M0\,V star (Leggeit 1992).

\begin{figure}
\resizebox{\hsize}{!}{\includegraphics{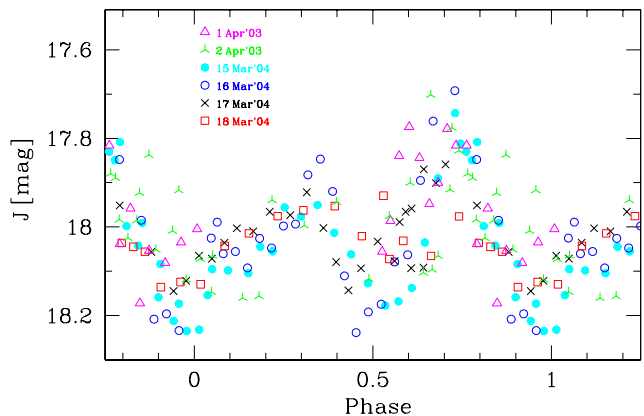}}
\caption{The combined $J$--band data from 2003 April and 2004 March folded on the orbital ephemeris, $ {\rm HJD\,Min} = 2451880.1086 + 0.1699339 \times E$  (Torres et al. 2004).}  
\label{lcfaz} \end{figure}

All our $J$ light curves (Fig.~\ref{lcmar04}) show modulation with a period of $\sim 0.17$ d, with minima roughly coinciding with the spectroscopic conjunctions of the secondary according to the ephemeris given by Torres et al. (2004). The combined light curve folded on this orbital ephemeris and composed from all our data (Fig.~\ref{lcfaz}), however, does not reveal a simple ellipsoidal modulation expected for the Roche--lobe filling star. Instead, both its shape and apparent asymmetries are almost identical with those shown by the superhump modulation found by Zurita et al. (2002a) in the optical light curves obtained during decline to quiescence, and  shown in the bottom panel of their figure 3.
This result is very surprising because if the near--IR light curves are dominated by the superhump modulation, it is implausible that the light curves separated in time by $\sim 1$ year would be in the same phase of the orbital period, with the minima occuring at the time of spectroscopic conjunctions.
Moreover, the quiescent near--IR luminosity of XTE 1118+480 is much higher than the quiescent  optical and UV luminosity (McClintock et al. 2004) which is inconsistent with the expected spectrum of an accretion disk emission.
Thus, most of the observed modulation presumably originates in the tidaly distorted secondary, although some contamination by the superhump modulation may also be present.

The light curves will be analysed in detail and possible interpretations of their complex shape will be discussed in a forthcoming paper. 

\subsection{Distance}

The presence of a Roche-lobe filling secondary provides an excellent opportunity to accurately estimate the distance $d$ to XTE\,J1118+480
For a Roche-lobe filling secondary, we have (Eggleton 1983):
\begin{equation}
\frac{R_2}{a}  = \frac{0.49 q^{2/3}}{0.6 q^{2/3} + \ln (1+q^{1/3})}
\end{equation} 
Adopting the mass ratio $q=M_2/M_\mathrm{X} = 0.037 \pm 0.007$ (Orosz 2001) and the secondary's radial velocity amplitude $K_2 = 709 \pm 7$ km/s (Torres et al. 2004) we estimate  the system separation $a \sin i = 2.47 \pm 0.03\, R_{\sun}$, the volume radius of the secondary $R_2/a = 0.154^{+0.008}_{-0.010}$ or $R_2 = (0.38 \pm 0.03)\sin^{-1} i\, R_{\sun}$. 
Using the Barnes-Evans relation (Cahn \cite{cahn80}),
\begin{equation}
F_\mathrm{K} = 4.2211 - 0.1 K - 0.5 \log s,
\end{equation}
with the $K$ surface brightness, $F_\mathrm{K} = 3.826$ (for $V-K \sim 3.6$ and [M/H]=--0.5; Beuermann et al. 1999), and $K=16.9 \pm 0.2$ (average value),
we estimate the secondary's angular diameter, $s=2.57 \times 10^{-3}\, \mathrm{mas}$, and the distance $d = (1.37 \pm 0.20) (R_2/0.38\, R_{\sun}) \sin^{-1} i\,  \mathrm{kpc}$. 
The surface brightness is a function of metallicity, lower metallicity will push the distance to lower values, the effect will, however, not exceed a few per cents. In particular, $d \sim 1.3$  for [M/H] $\sim  ~ -1$ (Frontera et al. 2001) and $d \sim 1.5$ kpc for  [M/H] = 0 (solar). 

The system is not eclipsing, so the inclination is $i \la 80^{\degr}$. Zurita et al. (2002a) estimate that $i$ lies in the range 71--82$^{\circ}$ based on analysis of the ellipsoidal light curve in the optical ($R$) range.  The presence of a strong orbital modulation in our $J$ light curves also  indicates high $i \sim 80^{\circ}$. So, most likely $\sin i \ga 0.95$, and  $d \approx 1.4 \pm 0.2$ kpc.

This estimate assumes that there is no accretion disc contribution to the near--IR magnitudes. 
Torres et al. (2004; see also Zurita et al. 2002a) found a constant $\sim 45\%$ contribution of light from the accretion disc to the quiescent optical red emission. Adopting the quiescent magnitude $R\sim18.9$--19 (Orosz 2005; Zurita et al. 2002b) and  assuming the frequency dependence of the disc emission,  $F_\nu \propto \nu^{1/3}$, we estimated the upper limit of such a contribution to the near--IR light of about 33\% at $J$ and 25\% at $K$. This will set the upper limit for the distance to XTE\,J1118+480 of $d \la 1.6$ kpc.
The present estimate is not in conflict with the previous less accurate estimates of $1.8 \pm 0.6$ kpc (McClintock et al. 2001), and $1.9\pm 0.4$ (Wagner  et al. 2001), respectively, based on optical spectral classification of the secondary.

\section{Conclusions}

The major results and conclusions of this paper can be summarised as follows:

\begin{description}
\item ({i})  We have detected for the first time  KV\,UMa, the optical component of XTE\,J1118+480, in the near infrared at quiescence.
\item ({ii}) The  near--IR magnitudes and colours are consistent with an K/M--type star which suggest that the secondary dominates in this spectral range.
\item ({iii}) The $J$ light show strong modulation with the orbital ephemeris of Torres et al. (2004). The light curves, however, show significant departures from a simple ellipsoidal modulation. 
The nature of the complex form of the $J$ curve is not clear and needs further studies. 
\item({iv}) The distance to XTE\,J1118+480 is $d \approx 1.4 \pm 0.2$ kpc assuming negligible contribution from the accretion disc, and the upper limit is $d \la 1.6$ kpc.
\end{description}

\subsection*{ACKNOWLEDGEMENTS} 
We thank Jorge Casares, Phil Charles and Andrzej Zdziarski, as well as the referee Jerome Orosz for the very helpful comments on this project. JM and DRG also thank Michael Pohlen for excellent introduction to near-IR photometry  
with the TCS/CAIN II, and for providing the IRAF CAIN data  
reduction package MPCAINRED.
This study was supported in part by the 
KBN Research Grants Nos  1\,P03D\,017\,27, PBZ-KBN-054/P03/2001, 1\,P03D\,018\,27, and the Spanish grant AYA 2001-1646.

\label{lastpage}

\end{document}